# Complex Dynamics of Bus, Tram and Elevator Delays in Transportation System

Takashi Nagatani

Department of Mechanical Engineering, Shizuoka University, Hamamatsu, Japan

## Article Outline

## Glossary

Nonlinear map model
    A nonlinear map is a recurrence relation presented by a nonlinear function. The nonlinear map model is a dynamic model of bus motion described by the nonlinear map.

Piecewise map model
    A recurrence relation presented by a continuous function comprised of segments. The piecewise map model is a dynamic model of bus motion described by the piecewise map.

Delayed map model
    Nonlinear map with a time lag. The delayed map model is a dynamic model of bus motion described by the delayed map where the delay of acceleration or deceleration is taken into account.

Extended circle map model
> The circle map is a one-dimensional map which maps a circle onto itself. A nonlinear map including sin function. The extended circle map model is a dynamic model of a bus with the periodic inflow at a transfer point

Combined map model
> The combined map is a nonlinear map combined with two nonlinear maps. The combined map model is a dynamic model of a single bus in which both effects of speed control and periodic inflow is taken into account.

Deterministic chaos
> The future behavior of dynamics involved no random elements does not make predictable. The behavior is known as deterministic chaos. The chaos occurs in the nonlinear map models of buses, trams, and elevators.

## Definition of the Subject

It is necessary and important to operate buses and trams on time. The bus schedule is closely related to the dynamic motion of buses. In this part, we introduce the nonlinear maps for describing the dynamics of shuttle buses in the transportation system. The complex motion of the buses is explained by the nonlinear-map models. The transportation system of shuttle buses without passing is similar to that of the trams. The transport of elevators is also similar to that of shuttle buses with freely passing. The complex dynamics of a single bus is described in terms of the piecewise map, the delayed map, the extended circle map and the combined map. The dynamics of a few buses is described by the model of freely passing buses, the model of no passing buses, and the model of increase or decrease of buses. The nonlinear-map models are useful to make an accurate estimate of the arrival time in the bus transportation.

## Introduction

The traffic science aims to discover the fundamental properties and laws and the traffic engineering aims at making the planning and implementation of road networks and control systems. Physics, other sciences, and technologies meet at the frontier area of interdisciplinary research. The traffic flow has been studied from a point of view of statistical physics and nonlinear dynamics [1-10]. Various models have been presented to understand the rich variety of physical phenomena exhibited by traffic. Analytical and numerical techniques have been applied to study these models. The concepts and techniques of physics can be applied to transportation systems. Specially, the traffic jams and congestion have been investigated. The vehicular traffic controlled by signals has been studied by some physical models [11-18]. Also, the dynamic models for multi-vehicle collisions have been proposed [19-21].

Some information of traffic has an important effect on the underlying dynamics. In the real traffic, advanced traveler information systems provide real-time information about the traffic conditions to road users by means of communication. Real-time information helps the individual road users to minimize their personal travel time. The dynamic models have been proposed for the route choice of traffic system with real-time information [22-29]. Despite the complexity of traffic, physical traffic theory is an example of a highly quantitative description for a living system.

In the public transportation system, it is important and necessary to make a schedule of public transport, for example bus schedule. It is well known that passengers using buses are served best when buses arrive at stations on time and there is no congestion. Frequently, buses are delayed or go faster in the bus transport system. Thus, it is hard to operate buses on time. Bus schedules are closely related to the dynamic motion of buses. The bus dynamics depends highly on the control method and the system. The arrival time of buses is not determined only by serving passengers but also by the speed control method of buses. The dynamics of trams and elevators is also similar to that of the bus transport system.

In the morning and evening peaks, the bus transport system exhibits severe congestion problems. The maximum rate of serving passengers increases with the number of buses. In managing the bus operation, the usual criterion for deciding the number of buses is that one should be able to transport everyone from the starting point to his destination within some period of time for the rush hour trips. Another criterion used in shuttle bus operation is that a passenger's waiting time should not exceed some specified value. It is very important to make an accurate estimate of the arrival time in the bus transportation. There also can be the same problems in the tram and elevator systems.

Until now, some models of the bus transport system have been studied. In the bus route model with many buses, it has been found that the bunching transition between a heterogeneously jammed phase and a homogeneous phase occurs with increasing density. Such dynamic models as the car-following model have been proposed to study the bunching transition [30-34]. Little is known about the travel time and delay of buses.

In shuttle bus systems, it is important to make correct bus schedules. It is necessary to calculate and derive the travel time and delay. The bus schedule must be determined by the dynamic motion of buses considering the inflow rate of passengers and the speed control method. By taking into account the speed control method and bus transport system, one should estimate the bus schedule. The dynamic model of the bus system is necessary to estimate the arrival time [35-42].

It is well known that the delay of acceleration and deceleration has the important effect on the jam formation in traffic flow. Many researchers have studied the delay effect for the vehicular traffic flow. However, the delay effect on bus schedule is unknown about the bus transportation system. It is important to study how the delay of speedup affects the bus dynamics and its schedule. There are some cases that the motion of buses is unstable if one retrieves the delay of buses. Buses show the periodic, quasi-periodic, and chaotic motions. It is difficult to make bus schedule due to the complex behavior.

In this part, we present the nonlinear-map models for the complex dynamics of buses, trams, and elevators. The bus schedule is derived from the nonlinear map model for the dynamic motion of a shuttle bus. We make an accurate estimate of the arrival time using the nonlinear map model. We show that both inflow rate and speedup control have an important effect on the dynamic transitions to complex motion occurring in the bus traffic. Also, we present the nonlinear map models for the transport system of a few buses. We discuss the relationship between the bus schedule and the dynamic motion.

## Model and similarity

We consider the dynamic model of shuttle bus system which mimics the service of buses shuttling between the origin and the destination (for example, an airport and a railway station). A bus carries the passengers from the origin to the destination. This shuttle bus model has the realistic background. We model the public transport system of shuttle buses as follows. Shuttle buses repeatedly go back and forth between the starting point (origin) and the destination. The starting point is the only position that passengers can take a bus. Passengers board a bus at the origin and then the bus starts from the origin. The bus moves toward the destination. When the bus arrives at the destination, all riding passengers leave the bus. After all passengers gets off the bus, the bus leaves the destination and returns to the origin. This process is repeated. This dynamics of the shuttle bus system can be described in terms of the nonlinear-map models.

Also, we consider the case in which a bus serves repeatedly between two terminals A and B. Passengers come into each terminal independently. Passengers can board a bus at both terminals. Each terminal is an origin and a destination. When the tram or airplane arrives at terminals periodically, the passengers come into bus terminals periodically. The dynamics of the bus system is described in terms of the extended circle map.

One can consider the bus system that a few buses shuttle between the origin and destination according as a few buses pass freely other buses. One compares this bus system with the elevator system that a few elevators shuttle between the entrance floor and the top floor. Passengers board the elevators at the entrance floor and passengers leave the elevators at the top floor. Then, the bus system has the similarity to elevators serving between the entrance floor and top floor. The bus dynamics is the same as that of elevators. Figure 1 shows the schematic illustration of two shuttle buses and two elevators [43-48].

Also, one can consider the bus system that a few buses shuttle between the origin and destination and each bus never pass over the other buses. One can compare this bus system with the tram system that a few trams make repeatedly round-trips between the origin and destination. Each tram cannot pass over other trams. Passengers board the tram at the origin, the tram moves toward the destination, and passengers leave the tram at the destination. Then, the tram system has the similarity to the bus system. The motion of trams has the same dynamics as the bus system [49,50]. In the ferry boat system, the dynamic motion of ferry boats is similar to the bus transport system and this dynamics can be described in terms of the nonlinear map model. [51]. In the air transport system, the dynamics of airplane is also described in the nonlinear map model [52].

One can consider the case that an operator controls the number of buses at the origin according to the number of passengers. We present the dynamic model of the bus system with increase or decrease of buses.

Here, we describe some dynamic models for the bus system since the elevator and tram systems are similar to the bus system.

### Nonlinear-map model

We consider a transport system of a single bus which carryies passengers from the origin to the destination repeatedly. We describe the dynamic model of the single bus system by the nonlinear map. We assume that all the passengers waiting at the origin can board the bus. New passengers arrive at the origin with inflow rate $\mu$ [persons/minute]. The arrival time at the origin and trip $n$ is defined by $t(n)$. So $\mu(t(n)-t(n-1))$ is the number of passengers that have arrived since the bus left the origin previously. The number of passengers boarding the bus at trip $n$ is expressed by

$$B(n) = \mu\bigl(t(n)-t(n-1)\bigr). \tag{1}$$

The passengers boarding the next coming bus are those waiting at the origin until the arrival time that the bus arrives again at the origin. After the bus arrives at the origin, the passengers coming into the origin cannot board the bus. In this case, the passengers getting on the bus are given by Eq. (1). If one takes into account capacity $F_{\max}$ of a bus, Eq. (1) changes

$$B(n) = \min[F_{\max}, \mu(t(n)-t(n-1))]. \tag{2}$$

We assume that the time which takes passengers to board the bus is proportional to the number of the passengers. The time is given by $\gamma B(n)$ where $\gamma$ is the time which takes one person to board a bus. Generally, as the number of passengers is increased, the time for passengers to board the bus is longer. Similarly, the time which takes passengers to leave the bus is given by $\beta B(n)$ where $\beta$ is the time which takes one person to leave a bus.

A bus driver can control the speed to retrieve the delay. A bus is accelerated or decelerated according to the tour time $\Delta t(n) (= t(n) - t(n-1))$. The speed depends on the tour time. The moving time of the bus is $2L/v(\Delta t)$ where $L$ is the distance between two terminals (the origin and the destination) and $v(\Delta t)$ is the speed of bus. The tour time equals the sum of these periods. The speed is determined by the tour time $\Delta t(n)$ at trip $n$. Then, the arrival time $t(n+1)$ of bus at the origin and trip $n+1$ is given by

$$t(n+1) = t(n) + (\gamma + \beta)\mu(t(n) - t(n-1)) + \frac{2L}{v(\Delta t(n))}. \tag{3}$$

The tour time $\Delta t(n+1)$ is given by

$$\Delta t(n+1) = (\gamma + \beta)\mu \Delta t(n) + \frac{2L}{v(\Delta t(n))}. \tag{4}$$

By dividing time by characteristic time $2L/v_0$ ($v_0$: reference speed), one obtains the map in dimensionless form

$$\Delta T(n+1) = \Gamma \Delta T(n) + \frac{1}{V(\Delta T(n))}, \tag{5}$$

where $\Delta T(n) = \Delta t(n)/(2L/v_0)$, $\Gamma = \mu(\gamma + \beta)$, and $V(\Delta T) = v(\Delta t)/v_0$.
The map (5) is the simple nonlinear map. The motion of bus is determined by map (5). The dependence of speed on the tour time is given by the control method.

### 1. Piecewise map model

If a tour time is higher than the critical value, a bus speeds up abruptly and moves with high speed $v_{high}$. Otherwise, the bus moves with normal speed $v_0$. Then, the map (5) is given by

$$\Delta T(n+1) = \Gamma \Delta T(n) + 1 \quad \text{if } \Delta T(n) \leq T_c,$$
$$\Delta T(n+1) = \Gamma \Delta T(n) + \frac{1}{V_{high}} \quad \text{if } \Delta T(n) > T_c, \tag{6}$$

where $V_{high} = v_{high}/v_0$ and $T_c = t_c v_0/2L$.

Eq. (6) is the linear piecewise map. The characteristic of the map is determined by the fixed point [53,54,55]. If the map has a stable fixed point, the tour time approaches the value of fixed point after a sufficiently large time even if the initial value of the tour time takes any value. We study the fixed point of the map. In Fig. 2, we display the piecewise map (6) for some values of dimensionless inflow rate $\Gamma$. Three curves are shown at $\Gamma = 0.1$, $\Gamma = 0.5$ and $\Gamma = 1.0$ where $v_{high}/v_0 = 2.0$ and $T_c = 1.5$. With increasing inflow rate $\Gamma$, the slope of the map increases. At $\Gamma = 0.1$, the map (6) has a stable fixed point since the slope at the fixed point is less than one. At $\Gamma = 0.5$, the map (6) does not have a stable fixed point but has an unstable fixed point because the absolute value of the slope at the fixed point is higher than one. At $\Gamma \geq 1$, the map (6) has no fixed point. If the map has a stable fixed point, the tour time approaches the value of the stable fixed point after sufficiently large time. The shuttle bus moves with the constant value of the tour time. If the map has an unstable fixed point, the tour time does not approach the constant value of the fixed point after sufficiently large time but it oscillates around the value of the fixed point. The shuttle bus moves periodically with trips. When the map does not have fixed points, the tour time diverges with increasing trip. The shuttle bus is late more and more with increasing trip because the passengers increase with increasing tour time.

The fixed point is given by

$$\text{stable fixed point } \Delta T_f = \frac{1}{1-\Gamma} \quad \text{for } 0 \leq \Gamma < \frac{T_c - 1}{T_c},$$

$$\text{unstable fixed point } \Delta T_f = T_c \quad \text{for } \frac{T_c - 1}{T_c} \leq \Gamma < \frac{2T_c - 1}{2T_c},$$

$$\text{stable fixed point } \Delta T_f = \frac{1}{2(1-\Gamma)} \quad \text{for } \frac{2T_c - 1}{2T_c} \leq \Gamma < 1, \tag{7}$$

where $v_{high}/v_0 = 2.0$.

We show the typical behavior of the tour time with varying the inflow rate. Figure 3 shows the plots of tour time $\Delta T(n)$ against trip $n$ at inflow rates $\Gamma = 0.40$ and $\Gamma = 0.63$ where $v_{high}/v_0 = 2.0$ and $T_c = 1.5$. The tour time oscillates with period 3 at $\Gamma = 0.40$ and period 4 at $\Gamma = 0.63$. Figure 4 shows the plot of the tour time against the inflow rate for $100 < n < 1000$. For $0 \leq \Gamma < 1/3$, the tour time displays a single curve and its value is consistent with the value of the stable fixed point. For $1/3 \leq \Gamma < 2/3$, the tour time takes multi values. If the tour time takes triple values, it oscillates with period 3. For $2/3 \leq \Gamma < 1$, the tour time displays a single curve and its value agrees with the value of the stable fixed point.

Generally, the boarding and leaving times are proportional to the power of the tour time. The linear piecewise map (6) is extended to the nonlinear piecewise map

$$\Delta T(n+1) = \Gamma \Delta T(n)^\alpha + 1 \qquad \text{if } \Delta T(n) \leq T_c,$$

$$\Delta T(n+1) = \Gamma \Delta T(n)^\alpha + \frac{1}{V_{high}} \qquad \text{if } \Delta T(n) > T_c, \qquad (8)$$

where $\alpha$ is a positive real number. In the extended version, the tour time displays more complex behavior than the map (6). The bus behaves not only periodically but also chaotically [53,54]. Figure 5 shows the plot of tour time $\Delta T(n)$ against inflow rate $\Gamma$ at $\alpha = 2.0$. In the region between points *a* and *b*, the tour time oscillates periodically. In the region higher than point *b*, the tour time displays chaotic or quasi-periodic behaviors.

The bus driver can control the speed of the bus by other methods. For example, the driver speeds up more and more when tour time $\Delta t$ approaches critical value $t_c$ and the bus speed reaches maximum speed $v_{max}$ after a while. We assume that the speed is controlled by the hyperbolic -tangent function. In this case, the map (6) changes to the hyperbolic-tangent map.

## 2. Delayed map model

If the speed is controlled with delay time $\tau$, the speed is described by $v(\Delta t(n-\tau))$. When $\tau = 1$, the speed is controlled by tour time $\Delta t(n-1)$ at trip *n*-1. The control of the speed is done per one round late. When $\tau = 2$, the speed is controlled by tour time $\Delta t(n-2)$ at trip *n*-2. In this case, the control of the speed is done per two rounds late. Then, the tour time for the bus system with delay $\tau$ is given by

$$\Delta T(n+1) = \Gamma \Delta T(n) + \frac{1}{V(\Delta T(n-\tau))}, \tag{9}$$

$$V(\Delta T(n-\tau)) = 1 + \frac{(v_{max}/v_0 - 1)}{2}\{1.0 + \tanh[A(\Delta T(n-\tau) - T_c)]\}, \tag{10}$$

Eq. (9) represents a kind of delayed map. Thus, we obtain the delayed map model for the tour time in the shuttle bus transportation [56]. The dynamics of a bus is described by the delayed map (9). The delayed map (9) is a $(1+\tau)$-dimensional map. The one-dimensional map with no delay changes to $(1+\tau)$-dimensional map (9) by introducing the delay of the speed control. The characteristics of the high dimensional map are very complex and it is difficult to derive the fixed point analytically. The delayed map has not studied but the delayed logistic map has been studied by Masoller and Rosso[57].

The effect of delay $\tau$ on the arrival time and bus schedule is studied. For comparison, the tour time is shown in the case of no delay $\tau = 0$. Figure 6 shows the plots of tour time $\Delta T(n)$ against inflow rate $\Gamma$ at $A = 5$ and $A = 8$ and $\tau = 0$ from $n$=100 to $n$=1000 where $v_{max}/v_0 = 2.0$ and $T_c = 1.5$. The tour time at $A = 5$ converges a single value in all the region and is indicated by a dashed line. At $A = 8$, the tour time converges a single value except for the region between points *a* and *b* and displays a single curve except for the region between points *a* and *b*. The tour time agrees with the fixed point except for the region between points *a* and *b*. However, the tour time does not converges a single value in the region between points *a* and *b* but has double values. This is due to the unstable fixed point in the region between points *a* and *b*. The tour time oscillates periodically with period 2.

Figure 7 shows the plot of tour time $\Delta T(n)$ against inflow rate $\Gamma$ at $A = 8$ and $\tau = 1$ from $n$=100 to $n$=1000 where $v_{max}/v_0 = 2.0$ and $T_c = 1.5$. The tour time converges a single value except for the region between points *a* and *b* and displays a single curve except for the region between points *a* and *b*. However, the tour time does not converges a single value in the region between points *a* and *b* but has many values. In the region between points *c* and *d*, the tour time takes four values and displays a periodic motion of period 4 with varying trip $n$. By introducing delay $\tau = 1$, the tour-time profile (Fig. 6) in the non-delayed map ($\tau = 0$) changes to that in Fig. 7.

Figure 8 shows the plot of tour time $\Delta T(n)$ against trip $n$ from $n$=100 to $n$=150 at $A$=8, $\tau = 1$ and $\Gamma = 0.4$ where $v_{\max}/v_0 = 2.0$ and $T_c = 1.5$. The tour time exhibits a roughly periodic motion with period 9 and varies irregularly in detail. Figure 9 shows the plot of tour time $\Delta T(n+1)$ against $\Delta T(n)$ from $n$=100 to $n$=1000 at $A$=8, $\tau = 1$ and $\Gamma = 0.4$ where $v_{\max}/v_0 = 2.0$ and $T_c = 1.5$. The return map in Fig. 9 corresponds to Fig. 8. The return map shows a smooth closed curve. The smooth closed curve is similar to that observed in the circle map [58]. The smooth closed curve is observed in the quasi-periodic motion. The dynamic motion of the bus displays the quasi-periodicity by introducing the speedup delay. Thus, the speedup delay has the important effect on the bus motion.

## 3. Extended circle map model

Generally, the arrival rate of the passengers is periodic or irregular. When a bus route is connected with other routes, the arrival rate of the passengers is periodic at the transfer point. We consider the bus system that the arrival rate of passengers is periodic at the terminals. In realistic situations, the demand (the arrival rate of the passengers) is not simply periodic but quasi-periodic or irregular, If the fluctuation of the demand is weak around the periodicity, the quasi-periodic demand can be taken into account as the superposition of the periodic part and noise part. The periodic passenger arrival rate simulation helps us understand the bus behavior for more realistic situations.

A single bus serves repeatedly between two terminals A and B. Passengers come into each terminal periodically. Passengers coming into terminal A are independent on those coming into terminal B. Passengers board the bus at terminal A, the bus moves toward terminal B, and all passengers get off the bus at terminal B when the bus arrives at terminal B. As soon as the bus is empty, passengers at terminal B board the bus, the bus moves toward terminal A, and all passengers get off the bus at terminal A when the bus arrives at terminal A. The process is repeated. In the previous model, the starting terminal is the only way to board the bus. Here we extend the previous model to take into account boarding at terminal B. The bus shuttles repeatedly between two terminals. When the tram or airplane arrives at terminals periodically, the passengers come into terminals periodically. Then, passengers board on the bus at terminals with periodic inflows.

The dynamic model for the shuttle bus with the periodic inflows is described by the nonlinear map [59-62]. Passengers come into terminal A (B) at the inflow rate with period $t_{s,A}$ ($t_{s,B}$). Define the number of passengers boarding the bus at terminal A (B) on trip $n$ by $B_A(n)$ ($B_B(n)$). So $\gamma B_{A(B)}(n)$ is the amount of time needed to board all the passengers at terminal A (B). The moving time of the bus between terminals A and B is $L/V$ where $L$ is the length between terminals A and B and $V$ is the mean speed of bus. The stopping time at terminal A (B) to leave off the passengers is $\beta B_{A(B)}(n)$ where parameter $\beta$ is the time which takes one passenger to leave the bus. The tour time equals the sum of these periods. When all passengers boarding the bus at terminal A get off the bus at terminal B, the bus becomes empty. As soon as the bus is empty, new passengers are ready to board the bus. At once, new passengers waiting at terminal B begin to board the bus. We refer the time as the ready time. Therefore, the ready time $t_B(n)$ of the bus at terminal B on trip $n$ is given by

$$t_B(n) = t_A(n) + \gamma B_A(n) + \frac{L}{V} + \beta B_A(n). \tag{11}$$

The ready time $t_A(n+1)$ of the bus at terminal A on trip $n+1$ is given by

$$t_A(n+1) = t_A(n) + \gamma B_A(n) + \frac{L}{V} + \beta B_A(n) + \gamma B_B(n) + \frac{L}{V} + \beta B_B(n). \tag{12}$$

Define $W_{A(B)}(n)$ as the number of passengers waiting at terminal A (B) just before the bus arrives at terminal A (B) on trip $n$. New passengers arrive at terminal A (B) at rate $\mu_{A(B)}(t)$. So $W_{A(B)}(n)$ is the number of passengers that have arrived since the previous bus left terminal A (B). $W_A(n)$ and $W_B(n)$ are expressed by

$$W_A(n) = \int_{t_A(n-1)}^{t_A(n)} \mu_A(t)dt, \tag{13}$$

$$W_B(n) = \int_{t_B(n-1)}^{t_B(n)} \mu_B(t)dt. \tag{14}$$

The periodic inflow rate is expressed by

$$\mu_{A(B)}(t) = \mu_{0,A(B)}\left\{1 + b_{A(B)}\cos(\frac{2\pi(t+\phi_{A(B)})}{t_{s,A(B)}})\right\}, \tag{15}$$

where $t_{s,A(B)}$ is the period of the inflow rate at terminal A (B), $\phi_{A(B)}$ is the phase shift of the inflow rate at terminal A (B) and $b_{A(B)}$ is the amplitude of fluctuating inflow rate at terminal A (B). Since $\mu_{A(B)}(t) \geq 0$, $0 \leq b_{A(B)} \leq 1$. By substituting Eqs. (13) and (14) into Eqs. (11) and (12) and dividing time by the characteristic time $2L/V$, one obtains the nonlinear maps of dimensionless ready time

$$\begin{aligned}T_A(n+1) = &T_A(n) + \Gamma_{0,A}(T_A(n) - T_A(n-1)) + \Gamma_{0,B}(T_B(n) - T_B(n-1))\\&+ \frac{\Gamma_{0,A}b_A T_{s,A}}{2\pi}\left\{\sin(\frac{2\pi(T_A(n)+\Phi_A)}{T_{s,A}}) - \sin(\frac{2\pi(T_A(n-1)+\Phi_A)}{T_{s,A}})\right\}\\&+ \frac{\Gamma_{0,B}b_B T_{s,B}}{2\pi}\left\{\sin(\frac{2\pi(T_B(n)+\Phi_B)}{T_{s,B}}) - \sin(\frac{2\pi(T_B(n-1)+\Phi_B)}{T_{s,B}})\right\} + 1\end{aligned} \tag{16}$$

$$\begin{aligned}T_B(n) = &T_A(n) + \Gamma_{0,A}(T_A(n) - T_A(n-1))\\&+ \frac{\Gamma_{0,A}b_A T_{s,A}}{2\pi}\left\{\sin(\frac{2\pi(T_A(n)+\Phi_A)}{T_{s,A}}) - \sin(\frac{2\pi(T_A(n-1)+\Phi_A)}{T_{s,A}})\right\} + 0.5\end{aligned} \tag{17}$$

where $T_{A(B)}(n) \equiv t_{A(B)}(n)V/2L$, $T_{s,A(B)} \equiv t_{s,A(B)}V/2L$, and $\Gamma_{0,A(B)} \equiv (\gamma+\beta)\mu_{0,A(B)}$. The nonlinear maps (16) and (17) of the ready times are the extended version of the circle map [58]. Thus, the dynamics of the shuttle bus is described in terms of the extended circle maps (16) and (17) for the shuttle bus with two periodic inflow rates. The dynamic property of the map is controlled by six parameters: loading parameters $\Gamma_{0,A}$, $\Gamma_{0,B}$, periods $T_{s,A}$, $T_{s,B}$, phases $\Phi_A$, and $\Phi_B$ when $b_A = b_B = 1$.

When $\Gamma_{0,B} = 0$, there are no passengers boarding the bus at terminal B, terminal A is the origin, terminal B is the destination, and passengers board only at terminal A and get off only at terminal B.

In the case of $T_{s,A} = T_{s,B}$ and $\Phi_A = \Phi_B = 0$, the inflow of passengers at terminal B synchronizes with that at terminal A. Both inflows of passengers have the same period and phase shift. Figure 10 shows the plot of tour time $\Delta T_A(n)$ versus loading parameter $\Gamma_{0,A}$ for arrival numbers from $n$=1000 to $n$=2000 for inflow periods $T_{s,A} = T_{s,B} = 2.7$. Black dots indicate the tour time for $\Gamma_{0,A} = \Gamma_{0,B}$. For comparison, we calculate the tour time in the case that passengers do not come into terminal B but come only into terminal A. Red dots indicate the tour time for $\Gamma_{0,B} = 0$. The tour time varies in a complex manner with the loading parameter. The tour time takes on multiple values except for a few points. By adding the inflow at terminal B to the bus transport, the tour time becomes about two times. The tour time increases with the loading parameter and diverges at $\Gamma_{0,A} = 0.5$. The value at the divergence point is half for the tour time of $\Gamma_{0,B} = 0$. The tour time drastically changes by adding the inflow at terminal B. The return map is shown for the shuttle-bus transport. Figure 11 shows the plots of $\Delta T_A(n+1)$ versus $\Delta T_A(n)$ for $T_{s,A} = T_{s,B} = 2.0$, $\Gamma_{0,A} = \Gamma_{0,B} = 0.31$ and $T_{s,A} = T_{s,B} = 2.7$, $\Gamma_{0,A} = \Gamma_{0,B} = 0.24$.

Return map on the upper side in Fig. 11 displays 19 points because the tour time varies periodically with period 19. Return map on the bottom in Fig. 11 shows the smooth closed curve. Therefore, the tour time varies quasi-periodically. Furthermore, the return map is calculated for various values of both loading parameter and inflow period. The return map displays either discrete points or smooth closed curve. The chaotic motion does not occur. Thus, the shuttle bus shows either periodic motion or quasi-periodic motion in the case of $T_{s,A} = T_{s,B}$ and $\Phi_A = \Phi_B = 0$.

In the case of $T_{s,A} \neq T_{s,B}$ and $\Phi_A = \Phi_B = 0$, passengers come periodically into terminal A with the period different from that of the inflow of passengers into terminal B and the phase shift of inflow A is the same as that of inflow B. Generally, the period of inflow at terminal A is different from that at terminal B because terminal A is connected with the bus route different from terminal B. The effect of the period difference on the bus motion is studied under the condition of the same phase shift. Figure 12 is the return map for $T_{s,A} = 2.7$, $T_{s,B} = 2.6$, and $\Gamma_{0,A} = \Gamma_{0,B} = 0.24$. By changing period $T_{s,B}$ from $T_{s,B} = 2.7$ to $T_{s,B} = 2.6$, the return map changes from the smooth closed curve in Fig. 11 to the breakup of the torus in Fig. 12. The quasi-periodicity in Fig. 11 changes to a chaotic motion. Thus, the quasi-periodicity of the bus changes to the chaotic motion by varying period $T_{s,B}$. The period difference induces the chaotic motion of the bus.

## 4. Combined map model

When passengers come periodically at origin and the bus driver controls the speed to retrieve the delay, the dynamics is described by the extended circle map combined with the hyperbolic-tangent map. The nonlinear map is called as the combined map [63,64]. The combined map of the dimensionless arrival time is given by

$$T(n+1) = T(n) + \Gamma_0(T(n) - T(n-1))$$
$$+ \frac{\Gamma_0 b T_s}{2\pi} \left\{ \sin(\frac{2\pi(T(n)+\Phi)}{T_s}) - \sin(\frac{2\pi(T(n-1)+\Phi)}{T_s}) \right\}. \quad (18)$$
$$+ \frac{1}{1 + \frac{1}{2}(\frac{v_{max}}{v_0} - 1)\{1 + \tanh[A(\Delta T(n) - T_c)]\}}$$

The map (18) consists of both map (9) and map (16) with $\Gamma_{0,B} = 0$. The shuttle bus shows the periodic, quasi-periodic, and chaotic motions by varying both periodic inflow and speed control. The chaotic motion does not occur only in the hyperbolic map or in the extended circle map but the chaotic motion occurs by combing the hyperbolic-tangent map with the extended circle map. Also, one is able to introduce the delay into the combined map. Then, the combined map (18) is more complex delayed combined map.

## Freely passing buses

We consider the case that a bus passes other buses freely when the bus reaches those. The dynamic model is presented for the fluctuation of riding passengers in a few shuttle buses [65-68]. Now we are looking at the case that passengers are only boarding at the origin and alighting at the destination. When the time headway between buses is shorter (longer), the awaiting passengers at the origin are fewer (more). The passengers boarding at the origin are few (many). In the result, the boarding and getting off times are shorter (longer). The bus may pass over the previous bus. Thus, buses run a neck- and –neck race with other buses. If the awaiting passengers are superior to the maximum capacity of the bus, the passengers corresponding to the capacity board the bus and the remaining passengers wait the next coming bus. Bus $i$ moves with speed $v_i$. Then, the dimensionless arrival time of bus $i$ at the origin is given by

$$T_i(m+1) = T_i(m) + \Gamma_0 B_i(m) + \frac{v_0}{v_i}, \quad (19)$$

with
$$W_i(m) = W_{i'}(m') - B_{i'}(m') + \Pi(T_i(m) - T_{i'}(m')) \quad (20)$$

and

$$B_i(m) = \min[F_{i,\max}, W_i(m)], \quad (21)$$

where $T_i(m) \equiv \frac{t_i(m)v_0}{2L}$, $\Gamma_0 \equiv \frac{(\gamma+\eta)v_0}{2L}$, and $\Pi \equiv \frac{\mu 2L}{v_0}$. Subscript $i'$ indicates the next coming bus after bus $i$.

The dynamics of the buses is described by the map (19) with (20) and (21). The map is iterated simultaneously for $M$ buses. The dynamical property of the map is controlled by four parameters: loading parameter $\Gamma(=\Gamma_0\Pi)$, the capacity $F_{i,\max}$, bus speed $v_i/v_0$, and bus number $M$. When perspective passengers increase, the value of loading parameter becomes high. For a few buses, the order of buses changes with trip $m$ and the time headway between buses changes from trip to trip because a bus passes other buses or is outstripped by other buses. As a result, the riding passengers also change from bus to bus. The buses exhibit a complex behavior.

The number of riding passengers with varying trips is calculated by iterating map (19) for the typical case of two buses. Figure 13 shows the plots of the numbers $B1(m)$ and $B2(m)$ of riding passengers on buses 1 and 2 against trip $m$ from trip $m=900$ to $m=950$ for $\Pi=40$ where $M=2$, $\Gamma_0=0.01$, $v_1=v_2=v_0$, and $F_{1,\max}=F_{2,\max}=50$. The dotted line indicates the maximum capacity $F_{1(or2),\max}$. The numbers $B1(m)$ and $B2(m)$ of riding passengers cannot be superior to $F_{1(or2),\max}$. For a low value of loading parameter, the numbers of riding passengers change irregularly from zero to a value lower than the maximum capacity $F_{1(or2),\max}$. For a high value of loading parameter, the numbers of riding passengers change irregularly from nonzero to the maximum capacity $F_{1(or2),\max}$. The buses carry frequently the full load of passengers.

The number of riding passengers changes by synchronizing the time headway. The time headway changes alternately from low to high values.

Both time headway and riding passengers vary almost periodically. However, these quantities exhibit an irregularity. When two buses arrive simultaneously at the origin, the time headway becomes zero. In the result, the riding passengers become zero because the waiting passengers are proportional to the time headway for no queuing at the origin.

Figure 14 shows the plot of the number $B1(m)$ of riding passengers on bus 1 against loading parameter $\Pi$ from sufficiently large trip $m$=900 to $m$=1000 without noises where $M$=2, $\Gamma = 0.01$, and $F_{1,\max} = F_{2,\max} = 50$. For low values of loading parameter $\Pi$, both number of riding passengers and time headway take the three values and their distributions exhibit the three peaks around the values. With increasing loading parameter $\Pi$, the localized distributions extend around the peaks and become the two extended distributions for $29.8 < \Pi < 47.4$. When loading parameter $\Pi$ is higher than 47.4, the two extended distribution breaks down three distributions again. If loading parameter $\Pi$ is higher than 66.5, the time headway keeps a constant value. For $\Pi < 66.5$, the time headway exhibits the chaotic motion. The distribution of the time headway exhibits the similar behavior to the number of riding passengers. When the loading parameter is higher than $\Pi = 44.0$, the number of riding passengers saturates some times at $F_{1(or2),\max}$. For $47.4 < \Pi < 66.5$, either bus 1 or bus 2 carries the full load of passengers. When loading parameter $\Pi$ is higher than 66.5, both numbers $B1(m)$ and $B2(m)$ saturate at the maximum capacity. At $\Pi_c = 66.5$, two buses becomes the full load. Therefore, the dynamical transition from the chaotic to regular motions at $\Pi_c = 66.5$ is induced by the full load buses.

In the case of freely passing buses, the behavior of buses exhibits a deterministic chaos even if there are no noises. The chaotic motion of buses is induced by the neck- and –neck race between buses. The dynamic behavior of buses is definitely different from that of the single bus.

### No passing buses

We consider the case that $N$ shuttle buses do not pass each other on the route. The dynamical model of $N$ shuttle buses is presented [69,70]. All buses are numbered from 1 (leading bus) to $N$ (latest bus) in regular order. The buses start in regular order of 1, 2, 3,⋯, $N$ at the starting point after perspective passengers board buses, move toward the destination, all currently boarding passengers leave the bus after the bus arrives at the destination, and return to the starting point. If a shuttle bus catches the other bus, its bus stops and restarts after delay time $T_{min}$ in order to keep the appropriate time headway. The bus does not change the speed except for stopping. The bus keeps constant speed until it stops. The new passengers arrive at the starting point at rate $\mu$. $\Delta T$ is the time headway between itself and its predecessor. When bus 1 does not catch bus $N$, the arrival time $T_1(n+1)$ of bus 1 at starting point at trip $n$+1 is given by

$$T_1(n+1) = T_1(n) + \Gamma\{T_1(n) - T_N(n-1)\} + 1 \quad \text{if} \quad T_1(n+1) \geq T_N(n). \quad (22)$$

If bus 1 catches bus $N$, bus 1 restarts after delay time $T_{min}$:

$$T_1(n+1) = T_N(n) + T_{min} \quad \text{if} \quad T_1(n+1) < T_N(n). \quad (23)$$

Similarly, the arrival time $T_i(n+1)$ of bus $i$ ($2 \leq i \leq N$) at starting point and trip $n$+1 is given by

$$T_i(n+1) = T_i(n) + \Gamma\{T_i(n) - T_{i-1}(n-1)\} + 1 \quad \text{if} \quad T_i(n+1) \geq T_{i-1}(n). \quad (24)$$

$$T_i(n+1) = T_{i-1}(n) + T_{min} \quad \text{if} \quad T_i(n+1) < T_{i-1}(n). \quad (25)$$

The bus motion depends on loading parameter $\Gamma$, delay time $T_{min}$, and number $N$ of buses. With increasing loading parameter $\Gamma$, the bus slows down since more passengers board the bus. With decreasing delay time $T_{min}$, the following bus becomes faster since less passengers board the bus. $N$ shuttle buses exhibit a complex behavior and dynamical transitions by varying values of the three parameters.

The behavior of two buses is derived by the use of iterates of Eqs. (22)-(25). Figure 15 shows the time evolutions of the tour time $\Delta T_1(n)$ and the time headway $\Delta T_{12}(n)$ for delay time $T_{min} = 0.8$ under the constant value $\Gamma = 0.6$ of loading parameter where $\Delta T_1(n) = T_1(n) - T_1(n-1)$ and $\Delta T_{12}(n) = T_2(n) - T_1(n)$. For $T_{min} = 0.8$, two buses oscillate irregularly. The oscillation exhibits a chaotic behavior. Fig. 16 shows the plot of tour time $\Delta T_1(n+1)$ against $\Delta T_1(n)$ over $n$=100-1000. This is the return map for Fig. 15. A circle indicates a value of tour time $\Delta T_1(n)$. The solid lines represent the trajectories of time-series plot over $n$=100-300. This shows that the bus behavior is a chaotic motion. The chaotic values of tour time lie on three straight lines with slopes 0.60, 1.60, and 2.0.

The dependence of tour time $\Delta T_1(n)$ on delay time $T_{min}$ is derived to study the dynamical transitions to the complex motions. Fig. 17 shows the plot of the values of tour time $\Delta T_1(n)$ against delay time $T_{min}$ for loading parameter $\Gamma = 0.6$. For a small value of $T_{min}$, two buses move with a constant value of tour time and bus 2 follows bus 1 with constant value $T_{min}$ of time headway. When delay time $T_{min}$ is less than the critical value $T_{min} = 0.48$, the tour time converges to the constant value. If delay time $T_{min}$ is higher than the critical value $T_{min} = 0.48$, the bus oscillates periodically. The dynamical transition from the regular motion to the periodic motion occurs at $T_{min} = 0.48$. At $T_{min} = 0.71$, the distinct transition from double periodic motion to the chaotic motion occurs. The value of Liapunov exponent changes from a negative value to a positive value at $T_{min} = 0.71$.

The phase diagram (region map) is derived for the distinct dynamical states and the dynamical transitions. Figure 18 shows the phase diagram in phase space $(T_{min}, \Gamma)$. The regular motion in which the tour time converges to a constant value is indicated by gray color. The periodic motion is represented by white color. The chaotic motion is indicated by the circle. With increasing loading parameter $\Gamma$, the region of regular motion enlarges and the region of periodic motion shrinks. The complex behavior of two shuttle buses depends highly on both loading parameter $\Gamma$ and delay time $T_{min}$.

In the case of three shuttle buses, the behavior similar to two buses is observed. The shuttle bus system with no passing is definitely different from that with passing freely. The passing has the important effect on the dynamic motion, traffic states, and dynamic transitions.

Increase or decrease of buses

We consider the case that an operator controls the number of buses at the origin according to the number of passengers [71-74]. 1~$N$ buses shuttle repeatedly between the starting point (origin) and the destination. Passengers arrive continuously at the origin with a rate. The number of buses increases one by one when the number of passengers is superior to the capacity, while the number of buses decreases one by one if the number of passengers is less than the capacity. For example, buses 1 and 2 start at the same time from the origin and move together. Buses 3-$N$ stop at the origin and wait for a chance. If the number of passengers is higher than the total capacity, bus 3 runs with buses 1 and 2. Buses 1-3 arrive at the destination at the same. Furthermore, if passengers increase, bus 4 runs with buses 1-3. When the number of passengers is less than the capacity, bus 2 stops at the origin and waits for a chance. Running buses return to the origin at the same time

　The increase (decrease) of buses is determined by the number of passengers waiting at the origin. However, it takes a time to arrange an additional bus. Therefore, a delay occurs when buses increase or decrease according to the number of passengers waiting at the origin. The delayed increase (decrease) of buses is taken into account in the bus transportation system. The delay is defined as $\tau$.
The capacity of a single bus is defined as $F_{max}$. In order to board all passengers waiting at the origin, $M(n-\tau)$ buses are necessary at trip $n$. The number of buses is given by

$$M(n-\tau) = 1 + \text{int}\left[\frac{B(n-\tau)}{F_{max}}\right], \tag{26}$$

where $B(n-\tau)$ is the number of passengers at trip $n-\tau$.
If passengers are distributed uniformly for $M$ buses at the origin, the number of passengers boarding a bus at trip $n$ is given by $B(n)/M(n-\tau)$. Then, the dimensionless arrival time $T(n+1)$ of buses at the origin at trip $n$+1 is given by

$$T(n+1) = T(n) + \frac{\Gamma(T(n) - T(n-1))}{M(n-\tau)} + 1. \tag{27}$$

The nonlinear map (27) is the delayed piecewise linear map. Delay $\tau$ occurs when buses increase or decrease according to the number of passengers waiting at the origin. When $\tau = 0$, the number of buses at trip $n$ is determined by the number of waiting passengers at trip $n$. For $\tau = 0$, Eq. (27) is a one-dimensional piecewise linear map. The fixed point is derived analytically. If $\tau = 1$, the increase or decrease of buses is one trip late. The delayed map (27) is a $(1+\tau)$-dimensional map. The one-dimensional map changes to $(1+\tau)$-dimensional map (27) by introducing the delay of the increase of buses. The characteristics of the high dimensional map are very complex and it is difficult to derive the fixed point analytically.

Figure 19 shows the piecewise linear map at $\Gamma = 2$. There is an unstable fixed point in region *c-d*. When the fixed point exists in the region *c-d*, two or three buses run in a complex manner because the fixed point is unstable.

Figure 20 shows the plot of dimensionless tour time $\Delta T(n)$ against dimensionless inflow rate $\Gamma$ for arrival numbers (trip) $n$ from $n$=100 to $n$=1000 at $\tau = 0$, and number limit $F_{max} = 50$. In the region between 0 and point *a*, a single bus runs with the constant value of tour time and the tour time agrees with the value of fixed point. In the region between points *a* and *b*, a bus or two buses run periodically and the tour time takes multiple values due to the periodic motion of buses. In the region between points *b* and *c*, two buses run with the constant value of tour time and the tour time agrees with the value of fixed point. In the region between points *c* and *d*, two or three buses run periodically and the tour time takes multiple values due to the periodic motion of buses. In the region between points *d* and *e*, three buses run with the constant value of tour time and the tour time agrees with the value of fixed point. In the region between points *e* and *f*, three or four buses run periodically and the tour time takes multiple values due to the periodic motion of buses. In the region between points *f* and *g*, four buses run with the constant value of tour time and the tour time agrees with the value of fixed point. In the region between points *g* and *4*, four or fix buses run periodically and the tour time takes multiple values due to the periodic motion of buses. Thus, the buses behave in a complex manner by increase or decrease of buses.

If there is a delay for arranging additional buses, the dynamic behavior of buses is very complex and shows the regular, periodic, and chaotic motions.

Here, we assumed that all buses start at the same time from the origin and move together. In this case, a bus does not pass over other buses. If passing is allowed, buses run a neck- and –neck race with other buses. The model of increase or decrease buses changes. The passing has an important effect on the bus dynamics by the neck- and –neck race.

## Future Directions

Here, we considered the single bus system on the single route and the bus system of a few buses on two routes. In the real transportation system, many buses run simultaneously on multiple routes. In a city, various transportation networks are formed in a complex manner. Men and women go to the destination by using buses, trams, and vehicles. People go from the starting point, through the transfer point, to the destination. One wishes to know how the trip time takes from the starting point, through the transfer point, to the destination. One has a think about where there are the transfer points. It is also important to select one proper transport system and route from various number of choices. Here, we presented some transport models of buses, trams, and elevators. The dynamic models of bus transport systems are simple compared to the real transportation system. It is necessary and important to extend the above models to be capable of the real transportation network. In the transportation network, it is important to make a route choice. It is necessary to extend the present models for the public transport model in the network. At the first step to the extension, one will be able to make a dynamic model in which the bus changes are taken into account. By accounting the bus changes at some transfer points, one can extend the nonlinear-map models to the coupled-map model.

In real traffic, the demand (the arrival rate of the passengers) varies irregularly and largely. When the demand is stochastic, weak stochasticity does not affect the qualitative features but strong stochasticity will have an important effect on the stability and dynamics of the system. It will necessary to investigate the effect of stochastic demand on the bus's motion.

## Bibliography

**Primary Literature**

**Books and Reviews**

Complex Dynamics of Bus, Tram and Elevator Delays in Transportation System, Figure 1

Schematic illustration for transport systems of two shuttle buses and two elevators. Two shuttle buses shuttle repeatedly between the origin and the destination. A shuttle bus passes freely over another bus. Two elevators serve from the entrance floor to the top floor. There is a similarity between transport systems of buses and elevators

Complex Dynamics of Bus, Tram and Elevator Delays in Transportation System, Figure 2

Three curves of piecewise map (6) at $\Gamma = 0.1$, $\Gamma = 0.5$ and $\Gamma = 1.0$. With increasing inflow rate $\Gamma$, the slope of the map increases. At $\Gamma = 0.1$, map (6) has a stable fixed point. At $\Gamma = 0.5$, map (6) has an unstable fixed point. At $\Gamma \geq 1$, the map (6) has no fixed point.

Complex Dynamics of Bus, Tram and Elevator Delays in Transportation System, Figure 3

Plots of tour time $\Delta T(n)$ against trip $n$ at inflow rates $\Gamma = 0.40$ and $\Gamma = 0.63$. The tour time oscillates with period 3 at $\Gamma = 0.40$ and period 4 at $\Gamma = 0.63$.

Complex Dynamics of Bus, Tram and Elevator Delays in Transportation System, Figure 4

Plot of the tour time against inflow rate $\Gamma$ for $100 < n < 1000$. For $0 \leq \Gamma < 1/3$ and $2/3 \leq \Gamma < 1$, the tour time displays a single curve and its value is consistent with the value of the stable fixed point. For $1/3 \leq \Gamma < 2/3$, the tour time takes multi values and displays the oscillation.

Complex Dynamics of Bus, Tram and Elevator Delays in Transportation System
, Figure 5

Plot of tour time $\Delta T(n)$ against inflow rate $\Gamma$ at $\alpha = 2.0$. In the region between points *a* and *b*, the tour time oscillates periodically. In the region higher than point *b*, the tour time displays chaotic or quasi-periodic behaviors.

Complex Dynamics of Bus, Tram and Elevator Delays in Transportation System
, Figure 6

Plots of tour time $\Delta T(n)$ against inflow rate $\Gamma$ at $A = 8$ and $A = 5$ where $\tau = 0$. The tour time at $A = 8$ converges a single value except for the region between points *a* and *b*. In the region between points *a* and *b*. the tour time oscillates periodically with period 2. The tour time at $A = 5$ converges a single value in all the region and is indicated by a dashed line.

Complex Dynamics of Bus, Tram and Elevator Delays in Transportation System
, Figure 7

Plot of tour time $\Delta T(n)$ against inflow rate $\Gamma$ at $A = 8$ and $\tau = 1$. The tour time displays a single curve except for the region between points *a* and *b*. The tour time does not converges a single value in the region between points *a* and *b* but has many values. In the region between points *c* and *d*, the tour time takes four values and displays a periodic motion of period 4 with varying trip *n*. By introducing delay $\tau = 1$, the tour-time profile (Fig. 6) in the non-delayed map ($\tau = 0$) changes to that in Fig. 7.

Complex Dynamics of Bus, Tram and Elevator Delays in Transportation System
, Figure 8

Plot of tour time $\Delta T(n)$ against trip *n* from *n*=50 to *n*=150 at *A*=8, $\tau = 1$

and $\Gamma = 0.4$. The tour time exhibits a roughly periodic motion with period 9 and varies irregularly in detail.

Complex Dynamics of Bus, Tram and Elevator Delays in Transportation System, Figure 9

The return map of the tour time in Figure 8. The return map draws a smooth closed curve.

Complex Dynamics of Bus, Tram and Elevator Delays in Transportation System, Figure 10

Plot of tour time $\Delta T_A(n)$ versus loading parameter $\Gamma_{0,A}$ for inflow periods $T_{s,A} = T_{s,B} = 2.7$. For comparison, the tour time is shown for $\Gamma_{0,B} = 0$. The tour time takes on multiple values except for a few points.

Complex Dynamics of Bus, Tram and Elevator Delays in Transportation System, Figure 11

Plot of $\Delta T_A(n+1)$ versus $\Delta T_A(n)$ at $T_{s,A} = T_{s,B} = 2.0$ and $\Gamma_{0,A} = \Gamma_{0,B} = 0.31$. The return map displays 19 points because the tour time varies periodically with period 19. Plot of $\Delta T_A(n+1)$ versus $\Delta T_A(n)$ for $T_{s,A} = T_{s,B} = 2.7$ and $\Gamma_{0,A} = \Gamma_{0,B} = 0.24$. The return map shows the smooth closed curve. The tour time varies quasi-periodically.

Complex Dynamics of Bus, Tram and Elevator Delays in Transportation System, Figure 12

Return map at $T_{s,B} = 2.7$, $T_{s,B} = 2.6$, and $\Gamma_{0,A} = \Gamma_{0,B} = 0.24$. By changing period $T_{s,B}$ from $T_{s,B} = 2.7$ to $T_{s,B} = 2.6$, the return map changes from the smooth closed curve in Fig. 11 to the breakup of the torus in Fig. 12. The quasi-periodicity in Fig. 11 changes to a chaotic motion.

Complex Dynamics of Bus, Tram and Elevator Delays in Transportation System, Figure 13

Plots of the numbers $B1(m)$ and $B2(m)$ of riding passengers on buses 1 and 2 against trip $m$ from trip $m = 900$ to $m = 950$ for $\Pi = 40$. The dotted line indicates the maximum capacity $F_{1(or2),\max}$. The numbers $B1(m)$ and $B2(m)$ of riding passengers are lower than $F_{1(or2),\max}$. The numbers of riding passengers vary irregularly.

Complex Dynamics of Bus, Tram and Elevator Delays in Transportation System, Figure 14

Plot of the number $B1(m)$ of riding passengers on bus 1 against loading parameter $\Pi$ from sufficiently large trip $m=900$ to $m=1000$ without noises. For low values of loading parameter $\Pi$, the number of riding passengers takes the three values and their distributions exhibit the three peaks around the values. With increasing loading parameter $\Pi$, the localized distributions extend around the peaks and become the two extended distributions for $29.8 < \Pi < 47.4$. When loading parameter $\Pi$ is higher than 47.4, the two extended distribution breaks down three distributions again. If loading parameter $\Pi$ is higher than 66.5, the time headway keeps a constant value. For $\Pi < 66.5$, the time headway exhibits the chaotic motion. When the loading parameter is higher than $\Pi = 44.0$, the number of riding passengers saturates some times at $F_{1(or2),\max}$. For $47.4 < \Pi < 66.5$, either bus 1 or bus 2 carries the full load of passengers. When loading parameter $\Pi$ is higher than 66.5, both numbers $B1(m)$ and $B2(m)$ saturate at the maximum capacity. At $\Pi_c = 66.5$, two buses becomes the full load. Therefore, the dynamical transition from the chaotic to regular motions at $\Pi_c = 66.5$ is induced by the full load buses.

Complex Dynamics of Bus, Tram and Elevator Delays in Transportation System, Figure 15

Time evolutions of tour time $\Delta T_1(n)$ and time headway $\Delta T_{12}(n)$ for delay time $T_{min} = 0.8$ under loading parameter $\Gamma = 0.6$. Two buses oscillate irregularly.

Complex Dynamics of Bus, Tram and Elevator Delays in Transportation System, Figure 16

The return map for Fig. 15. Plot of tour time $\Delta T_1(n+1)$ against $\Delta T_1(n)$. A circle indicates a value of tour time $\Delta T_1(n)$. The solid lines represent the trajectories of time-series plot over *n*=100-300.

Complex Dynamics of Bus, Tram and Elevator Delays in Transportation System, Figure 17

Plot of tour time $\Delta T_1(n)$ against delay time $T_{min}$ at loading parameter $\Gamma = 0.6$. For a small value of $T_{min}$, two buses move with a constant tour time and bus 2 follows bus 1 with constant value $T_{min}$ of time headway. When delay time $T_{min}$ is less than the critical value $T_{min} = 0.48$, the tour time converges to the constant value. If delay time $T_{min}$ is higher than the critical value $T_{min} = 0.48$, the bus oscillates periodically. The dynamical transition from the regular motion to the periodic motion occurs at $T_{min} = 0.48$. At $T_{min} = 0.71$, the distinct transition from double periodic motion to the chaotic motion occurs.

Complex Dynamics of Bus, Tram and Elevator Delays in Transportation System, Figure 18

Phase diagram in phase space $(T_{min}, \Gamma)$. The regular motion is indicated by gray color. The periodic motion is represented by white color. The chaotic motion is indicated by the circle. With increasing loading parameter $\Gamma$, the region of regular motion enlarges and the region of periodic motion shrinks. The complex behavior of two shuttle buses depends highly on both loading parameter $\Gamma$ and delay time $T_{\min}$.

Complex Dynamics of Bus, Tram and Elevator Delays in Transportation System , Figure 19

Piecewise linear map at $\Gamma = 2$. There is an unstable fixed point in region *c-d*. When the fixed point exists in the region *c-d*, two or three buses run in a complex manner because the fixed point is unstable.

Complex Dynamics of Bus, Tram and Elevator Delays in Transportation System , Figure 20

Plot of tour time $\Delta T(n)$ against inflow rate $\Gamma$ at $\tau = 0$ and capacity $F_{\max} = 50$. In the region between 0 and point *a*, a single bus runs with the constant value of tour time and the tour time agrees with the value of fixed point. In the region between points *a* and *b*, a bus or two buses run periodically and the tour time takes multiple values due to the periodic motion of buses. In the region between points *b* and *c*, two buses run with the constant value of tour time and the tour time agrees with the value of fixed point. In the region between points *c* and *d*, two or three buses run periodically and the tour time takes multiple values due to the periodic motion of buses.

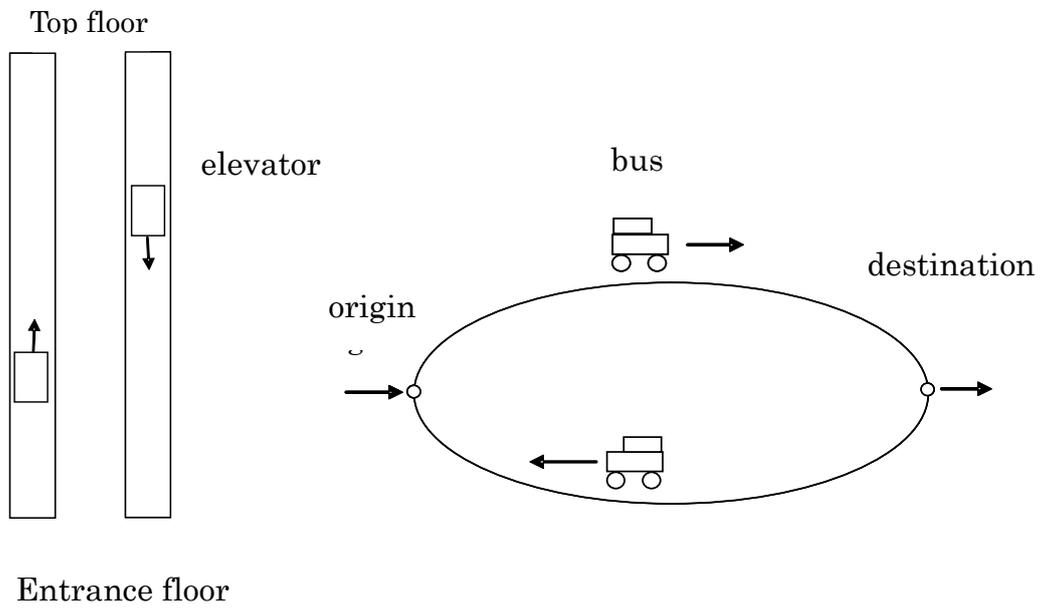

Figure 1

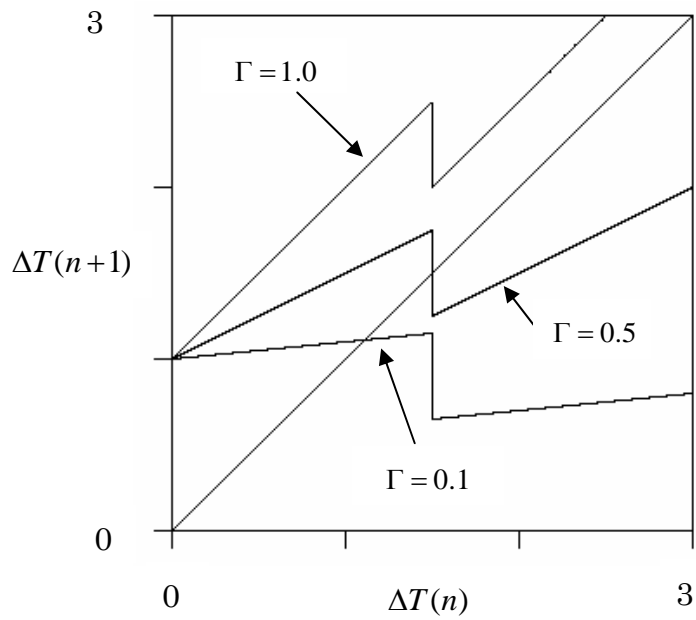

Figure 2

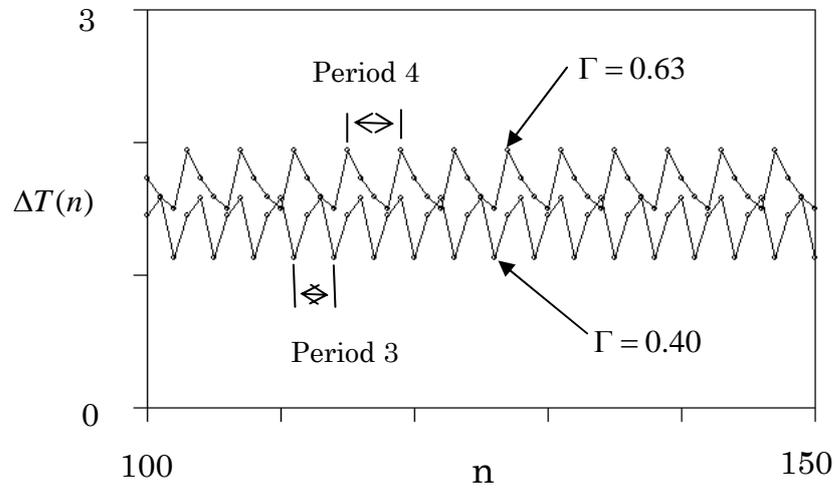

Figure 3

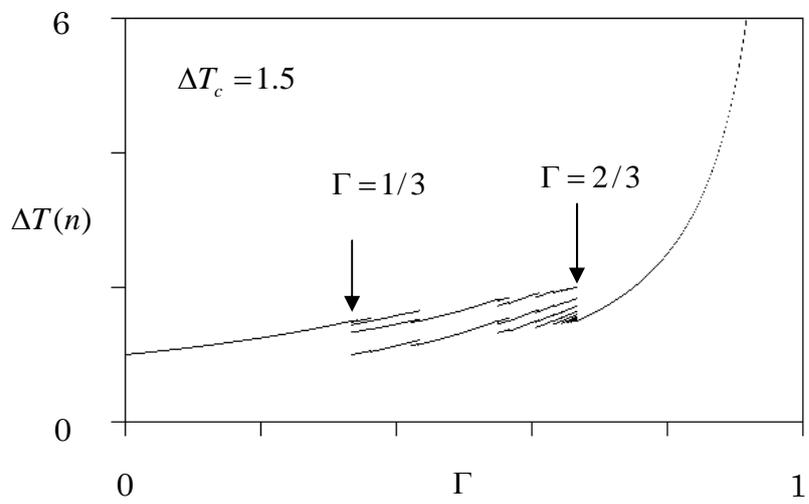

Figure 4

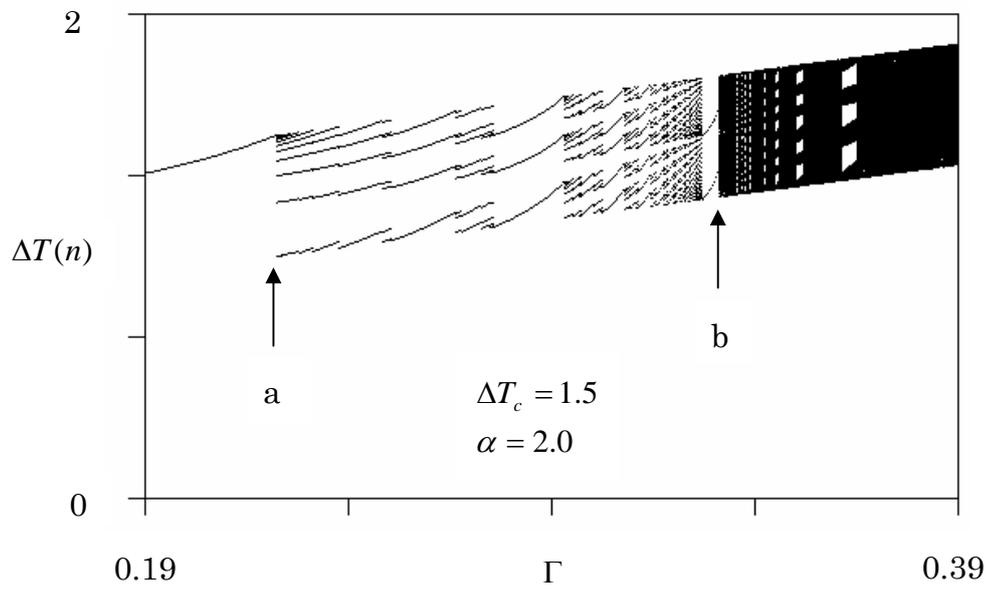

Figure 5.

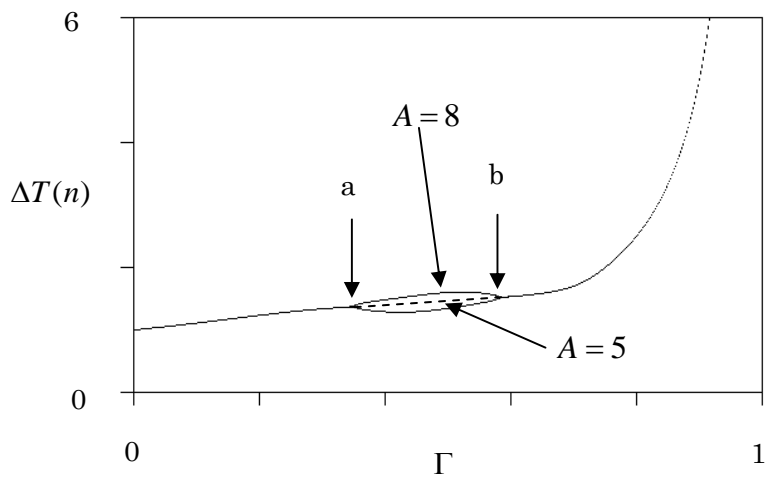

Figure 6.

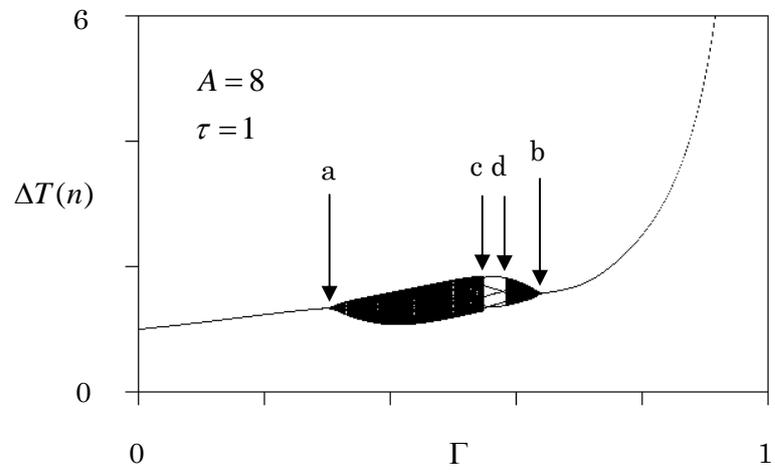

Figure 7

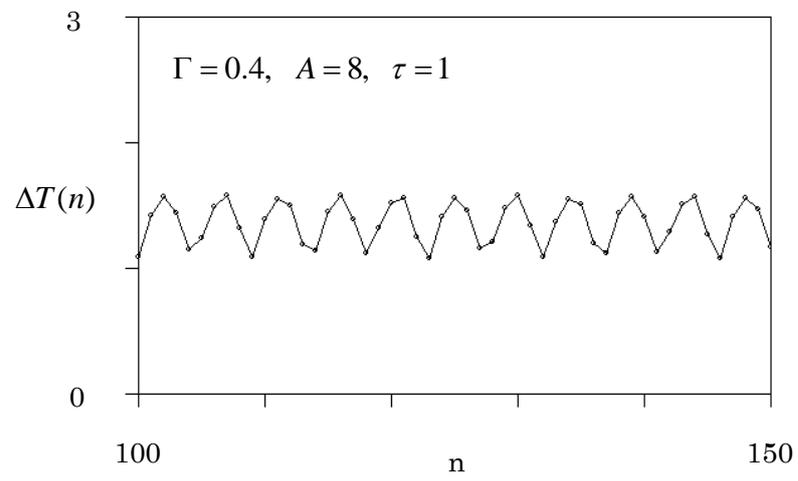

Figure 8

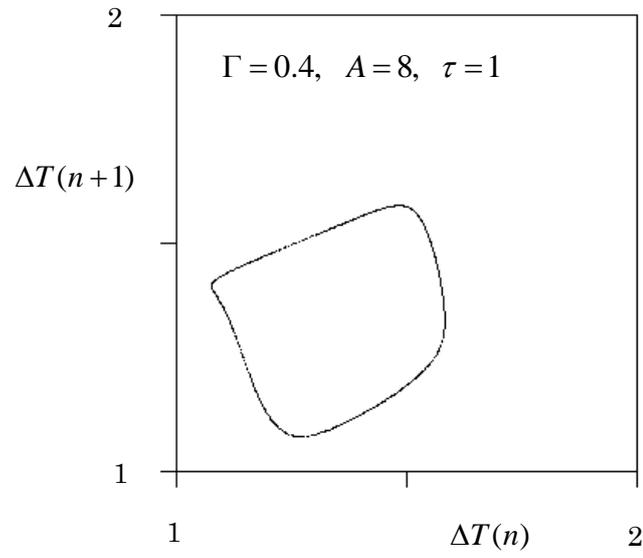

Figure 9

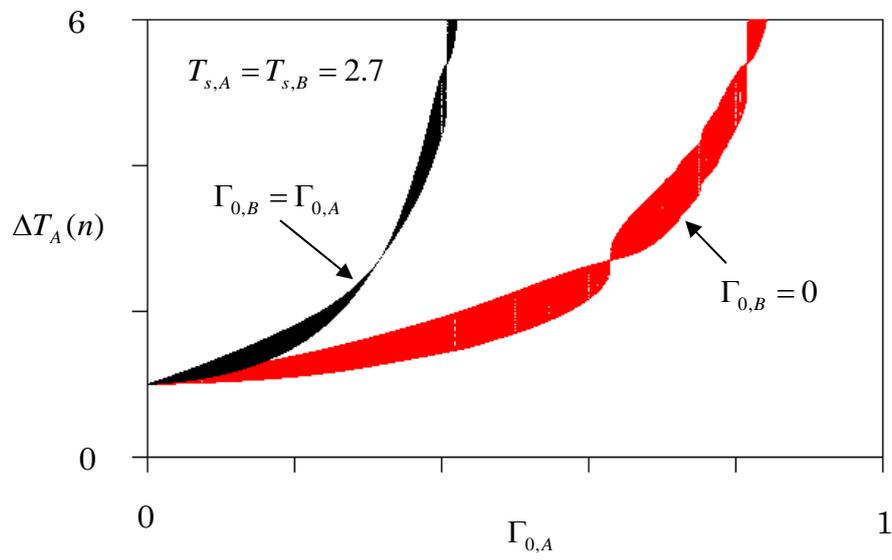

Figure 10

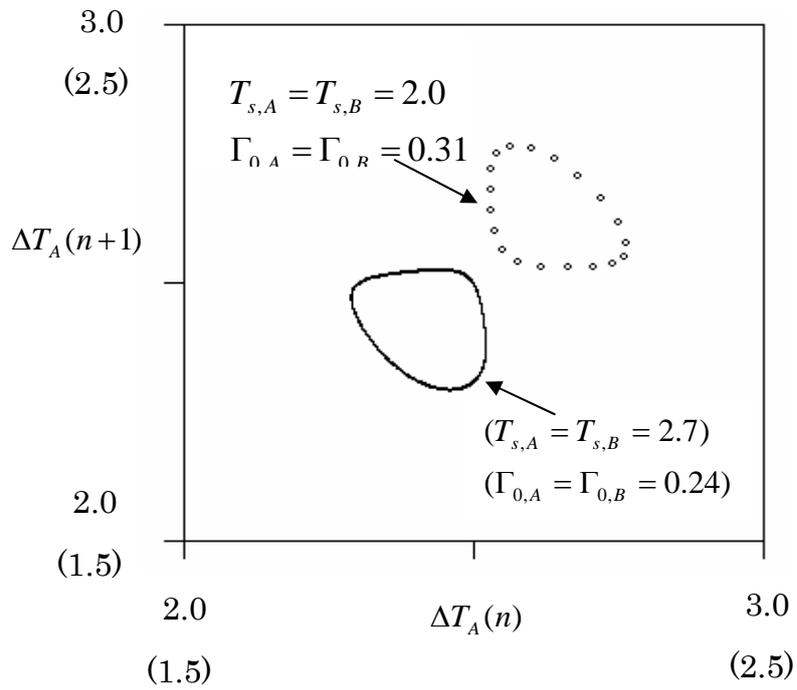

Figure 11

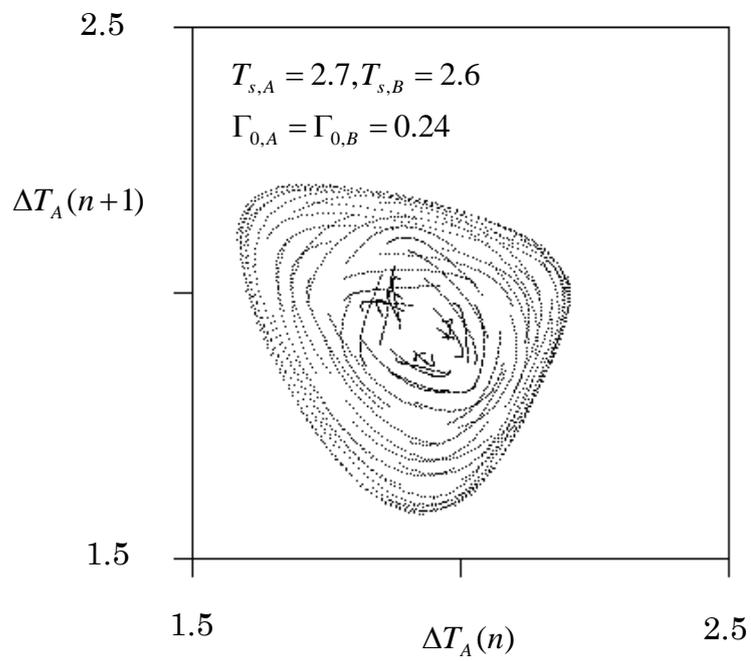

Figure 12

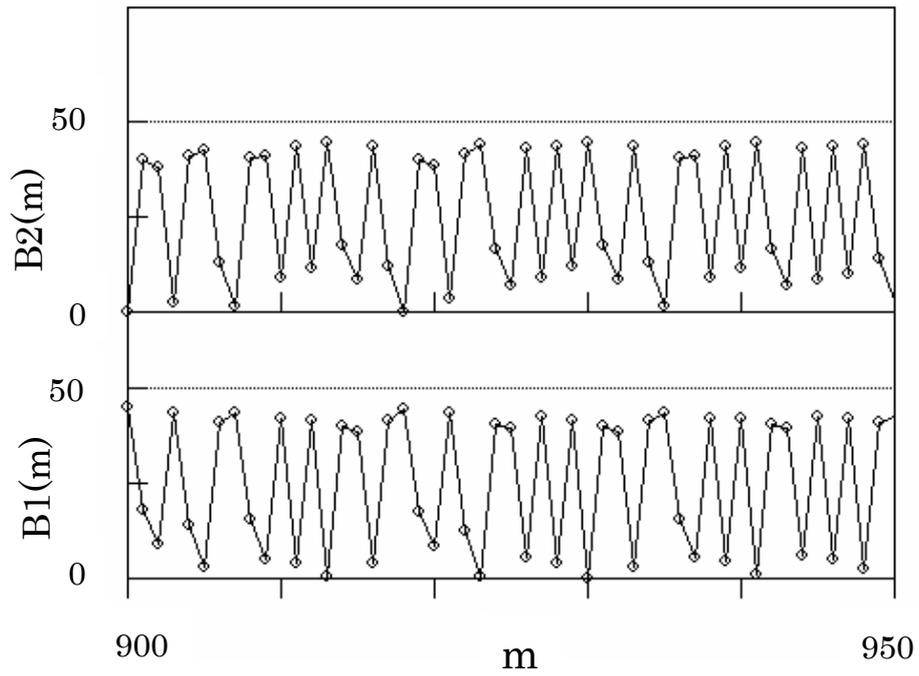

Figure 13

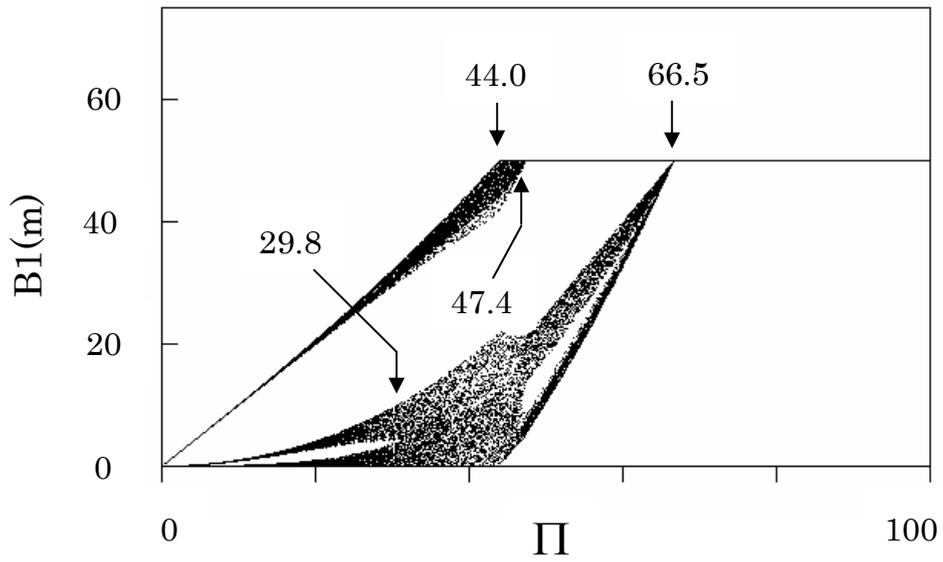

Figure 14

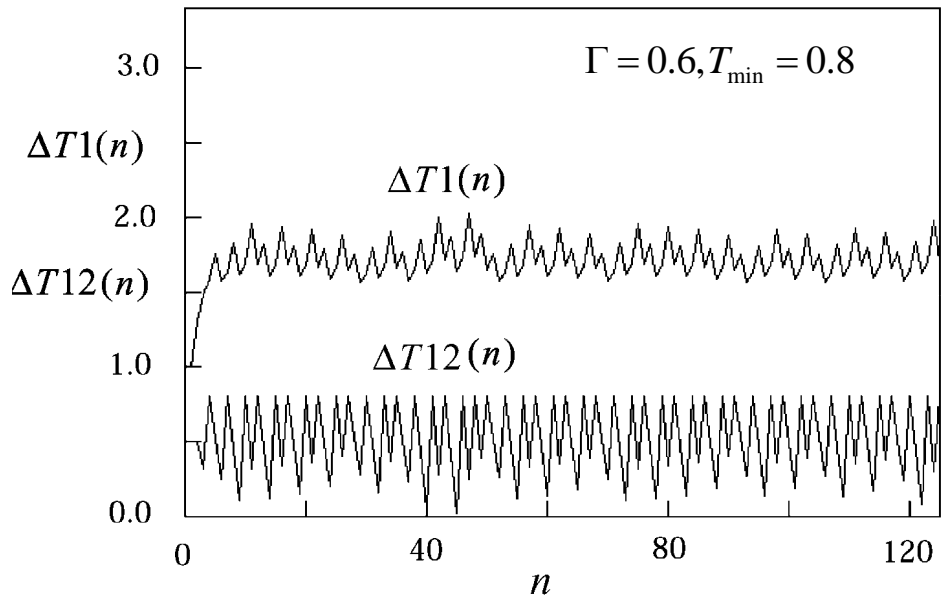

Figure 15

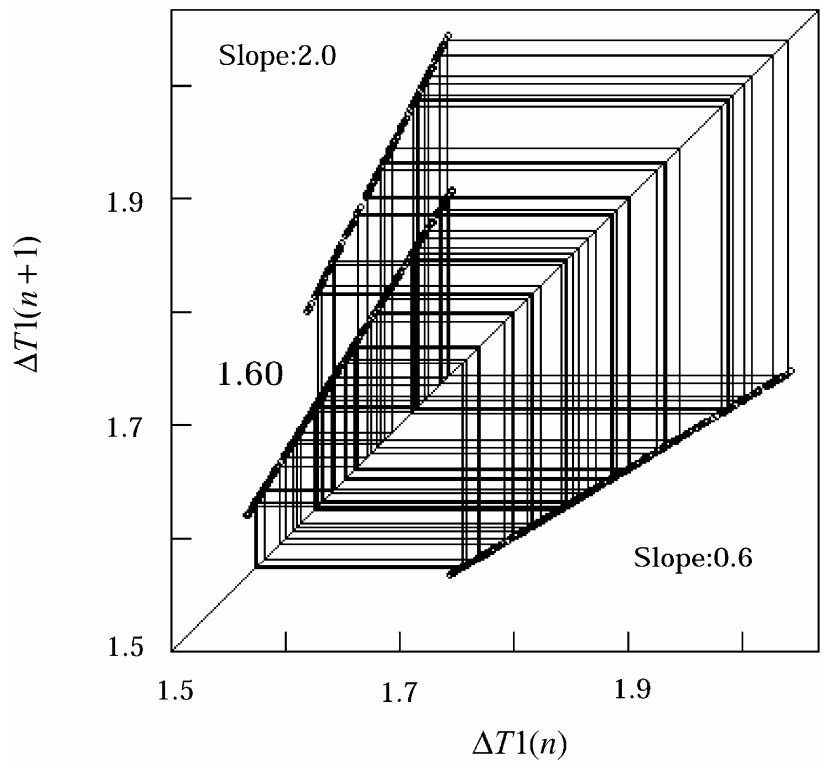

Figure 16

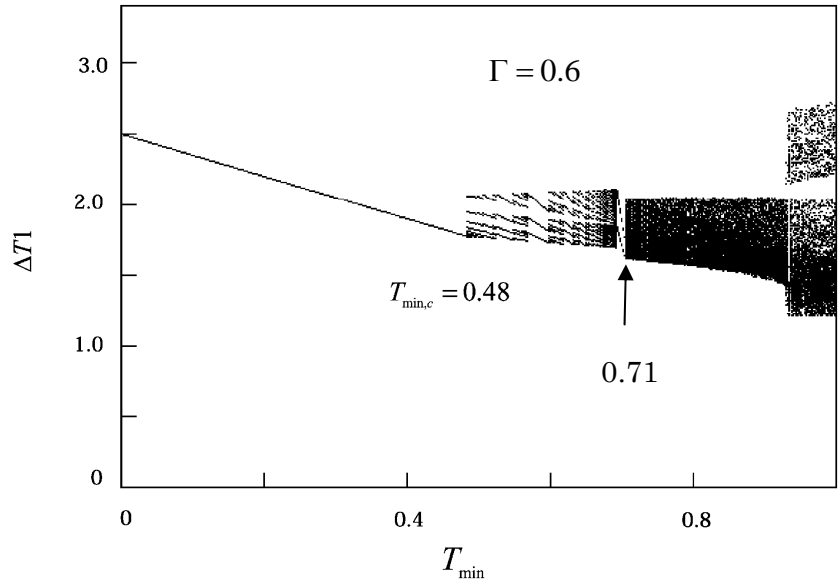

Figure 17

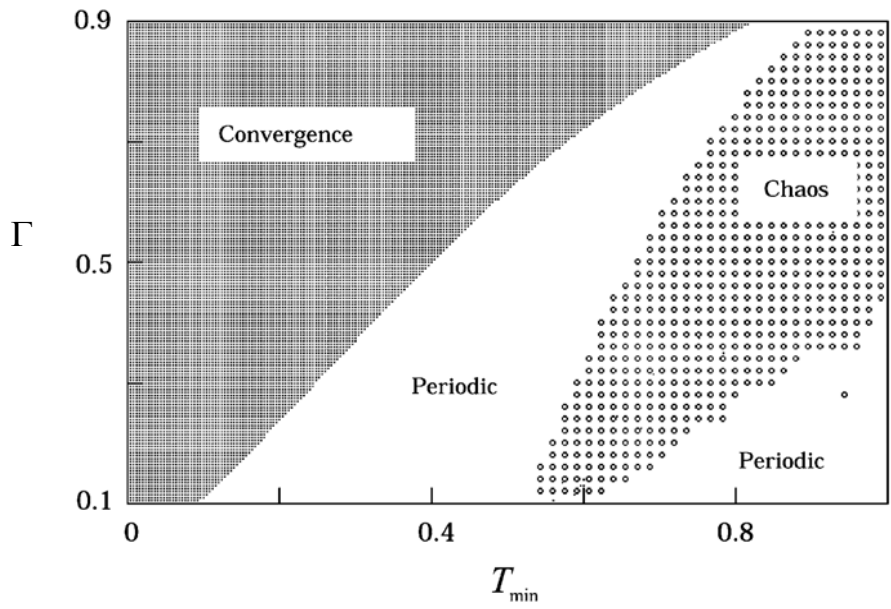

Figure 18

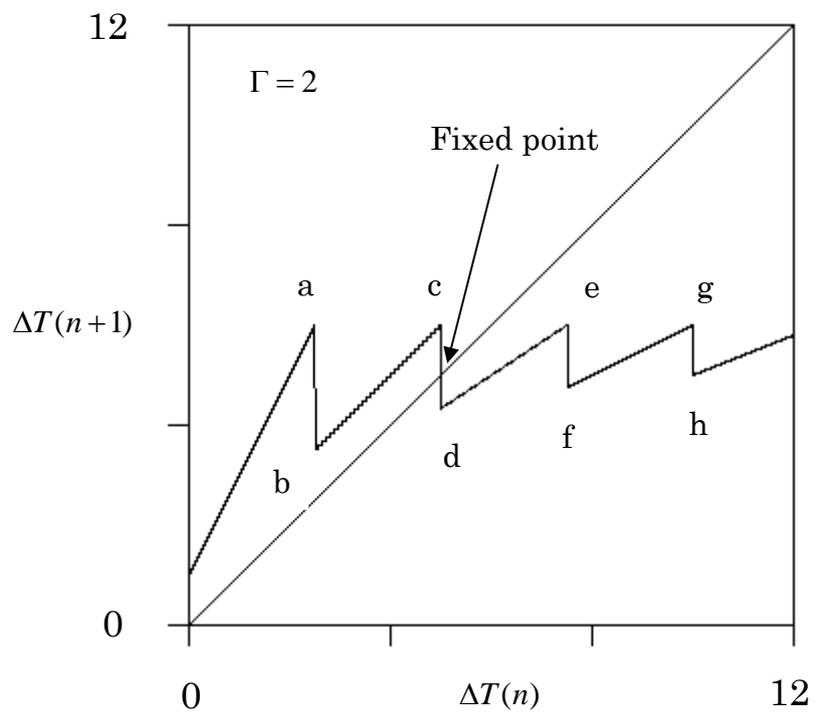

Figure 19

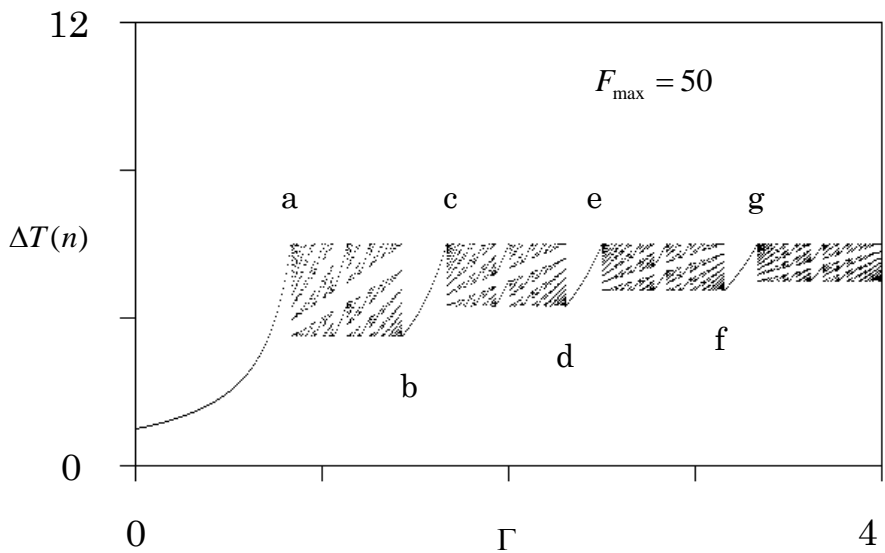

Figure 20